\documentclass[conference]{IEEEtran}
\IEEEoverridecommandlockouts
\usepackage{cite}
\usepackage{amsmath,amssymb,amsfonts}
\usepackage{dblfloatfix}
\usepackage{algorithmic}
\usepackage{graphicx}
\usepackage{hyperref}
\usepackage{textcomp}
\usepackage{caption}
\usepackage{subcaption}
\usepackage{xcolor}
\usepackage{comment}
\usepackage[export]{adjustbox} 
\usepackage{multirow}
\usepackage{verbatim}
\usepackage{listings}
\usepackage[normalem]{ulem}
\usepackage[ruled,vlined, linesnumbered]{algorithm2e}
\SetKwRepeat{Do}{do}{while}
\usepackage[colorinlistoftodos]{todonotes}
\def\BibTeX{{\rm B\kern-.05em{\sc i\kern-.025em b}\kern-.08em
    T\kern-.1667em\lower.7ex\hbox{E}\kern-.125emX}}
    
\newcommand{\method}{ACTA }

\begin{document}

\title{Automated Performance Testing Based on \\Active Deep Learning}

\author{\IEEEauthorblockN{Ali Sedaghatbaf}
\IEEEauthorblockA{\textit{RISE Research Institutes of Sweden} \\
Västerås, Sweden \\
ali.sedaghatbaf@ri.se}
\and
\IEEEauthorblockN{Mahshid Helali Moghadam}
\IEEEauthorblockA{\textit{RISE Research Institutes of Sweden} \\
Västerås, Sweden \\
mahshid.helali.moghadam@ri.se}
\and
\IEEEauthorblockN{Mehrdad Saadatmand}
\IEEEauthorblockA{\textit{RISE Research Institutes of Sweden} \\
Västerås, Sweden \\
mehrdad.saadatmand@ri.se}
}

\maketitle

\begin{abstract}
Generating tests that can reveal performance issues in large and complex software systems within a reasonable amount of time is a challenging task. On one hand, there are numerous combinations of input data values to explore. On the other hand, we have a limited test budget to execute tests. What makes this task even more difficult is the lack of access to source code and the internal details of these systems. In this paper, we present an automated test generation method called \method for black-box performance testing. \method is based on \textit{active learning}, which means that it does not require a large set of historical test data to learn about the performance characteristics of the system under test. Instead, it dynamically chooses the tests to execute using uncertainty sampling. \method relies on a conditional variant of \textit{generative adversarial networks}, and facilitates specifying performance requirements in terms of conditions and generating tests that address those conditions.
We have evaluated \method on a benchmark web application, and the experimental results indicate that this method is comparable with random testing, and two other machine learning methods, i.e. PerfXRL and DN.
\end{abstract}

\begin{IEEEkeywords}
Performance testing, automated test generation, active learning, conditional generative adversarial networks. 
\end{IEEEkeywords}

\section{Introduction}\label{sec:intro}
The widespread usage of software applications especially in critical environments makes their quality assurance of paramount importance. The ever-increasing complexity and changes in today's software systems make this task even more challenging. Global reports indicate that current \textit{verification and validation} (V\&V) methods are not effective in preventing the irrecoverable damages and daunting effects that software failures may have on the enterprises around the world. In 2018, poor software quality led to around \$3T loss only across the US \cite{news18}. Also, a recent study by Cambridge Judge Business school revealed that the software market spends \$61B on software failures \cite{news20}.

Performance is one of the fundamental software quality attributes, whose assessment is crucial to guarantee that a deployed software functions correctly \cite{ferme2017towards}, and it is perceived as \textit{the new correctness} in many practical deployment scenarios \cite{incerto2020inferring}. Performance issues may lead to several problems from degradation of performance indicators (e.g. response time or throughput) to complete system failures. However, finding and fixing performance issues is a difficult and challenging task \cite{porres2020automatic}. 
Several modeling and testing approaches have been proposed by researchers to analyze the performance of software systems. The goal of model-based approaches \cite{arcaini2020automated,koziolek2010performance, balsamo2004model} is to mimic the system behavior using a mathematical or simulation technique, and analyze the effects of various workloads, operational profiles, and service demands on the system performance through performance indicators. However, in performance testing, some test scripts are executed on the \textit{system under test} (SUT) to find potential scenarios which lead to the violation of performance requirements. 

Generating appropriate tests that reveal performance issues is a challenging task in performance testing. In fact, we usually have a large input space to explore, and trying all possible input value combinations for test  generation is impractical and laborious and even infeasible in many cases. Furthermore, regarding the time limitation in delivering efficient software products, exploring a large input space in a relatively short period of time is cumbersome. Test generation is even more challenging when it comes to black-box testing where we do not have access to the internal structure and dynamics of the SUT, and public interfaces are the only way to interact with the SUT and learn about its performance characteristics. 

Practically, a few input values trigger performance issues \cite{ahmad2019exploratory}. However, finding those values is mostly manual, intellectually intensive and laborious \cite{shen2015automating}. In recent years, researchers have proposed search-based profilers \cite{shen2015automating,luo2016mining}, fuzzers \cite{lemieux2018perffuzz}, symbolic execution methods \cite{chen2016generating,chattopadhyay2017directed,saumya2019xstressor}, and machine learning (ML) methods \cite{luo2017forepost,porres2020automatic,koo2019pyse} to find those input values automatically and in a cost-effective way. However, most of these methods (e.g. \cite{shen2015automating,luo2016mining,chen2016generating,saumya2019xstressor}) are suitable for white-box settings, where we have access to the source code of the SUT, which is not true in most cases \cite{ahmad2019exploratory}. Methods based on symbolic execution do not scale to systems with a large set of input data, since the number of paths to search grows exponentially with respect to the input size \cite{koo2019pyse}. 

Among the relate work, DN \cite{porres2020automatic}) and PerfXRL \cite{ahmad2019exploratory} are the most related to this paper. DN is a supervised (deep) learning method for test generation, which is well-suited for black-box settings. This method is passive, which means that it requires to execute numerous (random) tests to learn about the SUT and achieve a high level of accuracy. Nevertheless, if we let the ML model itself choose the tests that need to be executed (i.e. active learning (AL) \cite{settles2009active}), we can achieve the same (or even better) outcomes with less testing effort. Furthermore, the ML model used in this method is discriminative rather than being generative. In fact, given a set of random tests, this model can distinguish between positive and negative tests. So, after training the model, we still need to generate several random tests and use the model to find some positive ones. By \textit{positive} we mean tests that can impose resource-intensive computations on the SUT and degrade its performance. As a reinforcement learning method, PerfXRL supports only discrete input variables, whereas most systems have continuous input variables (e.g. speed and position) whose discretization may lead to state space explosion.

In this paper, we consider automatic and efficient test generation as a semi-supervised learning problem where the goal is to learn combinations of input values that can be used to generate positive performance tests. We assume a black-box setting where we have access to (1) the set of input parameters and the range of possible values for each, and (2) a test driver for test execution, which is supplied with a test oracle to determine whether the executed tes was positive or negative. To solve this problem, we propose an active deep learning (ADL) method called \textit{ACtive learning for Test Automation} (ACTA), which relies on \textit{conditional generative adversarial networks} (CGANs) \cite{mirza2014conditional} for learning about the performance of the SUT and generating positive tests for it. GANs \cite{goodfellow2014generative} are deep neural networks that can be trained to generate new plausible examples for a given dataset, and CGANs are a variant of GANs that facilitate generating samples conforming to a set of conditions. In \method we take advantage of CGANs to generate positive tests that address the performance requirements specified as conditions. \method uses uncertainty sampling \cite{settles2009active} to choose tests whose execution reveals information about the SUT that is more useful for updating the CGAN model. Our experimental results confirm the good performance of \method in comparison to DN, PerfXRL and random testing. In summary, the contributions of this paper include the following:
\begin{itemize}
    \item \method as a fully-automated and ADL-based method for cost-effective performance test generation is introduced,
    \item the architecture of the CGAN and the AL sampling scenario are elaborated,
    \item the algorithms for training the ADL model and generating positive tests are explained,
    \item the practicality of \method is illustrated through a case study on a benchmark web application, and
    \item the ease of adapting \method to DevOps is demonstrated.
\end{itemize}

The rest of the paper is organized as follows. 
Some background information about AL and CGANs is provided in Section \ref{sec:backgrnd}. An overview of \method is presented in Section \ref{sec:method}, and the experimental results are detailed in Section \ref{sec:exp}. The adaptability of \method to DevOps is demonstrated in Section \ref{sec:discuss}. Section \ref{sec:related} is dedicated to discussion about the related work, and, finally some concluding remarks and future directions are pointed out in Section \ref{sec:conclusion}.

\section{Background}\label{sec:backgrnd}
\subsection{Active Learning}
Active Learning (AL) \cite{settles2009active} is a sub-field of semi-supervised learning, where a small amount of unlabeled data is selected for labeling in an iterative and cost-effective manner. In fact, AL is suitable for situations that data labeling is expensive or difficult. AL algorithms typically involve a query scenario and use some query strategy to identify the unlabeled data that are more informative if labeled. Query scenarios include membership query synthesis, pool-based sampling, and stream-based sampling \cite{settles2009active}. In the first scenario, new data instances are generated from the uncertainty region of the classifier. In the other scenarios, instances are sampled either from a data pool or a data stream. Query strategies include uncertainty sampling, query-by-committee, variance reduction, etc \cite{settles2009active}. 

In this paper we follow the pool-based sampling scenario and use uncertainty sampling as the query strategy. Uncertainty sampling is among the simplest and most popular query strategies where an active learner queries the instances about which it is least certain how to label. To detect the instances with least certainty, 
the \textit{least confidence} (LC) measure is used, which can be formulated as follows \cite{settles2009active}:

\begin{equation}\label{eq:lc}
    x_{LC} = \underset{x}{\operatorname{argmax}}\, 1 - P_\theta(y^*|x) 
\end{equation}

\noindent where $x_{LC}$ is the instance with the least confidence about its class, and $y^*$ is the highest posterior probability predicted by model $\theta$. Margin sampling \cite{scheffer2001active} and entropy \cite{shannon2001mathematical} are other uncertain sampling strategies typically used in AL methods.

\subsection{Conditional Generative Adversarial Networks}\label{sec:CGAN}
Generative Adversarial Networks (GANs) \cite{goodfellow2014generative} are deep neural networks that can be trained to generate new plausible examples for a given dataset. Each GAN model consists of two neural networks called \textit{generator} and \textit{discriminator} that contest with each other in a zero-sum game. Generator synthesizes new candidates with the goal of fooling the discriminator such that it thinks the synthesized candidates are real ones, and discriminator evaluates each input candidate with the goal of detecting whether it's real or synthetic. After training, discriminator is discarded and generator can be used to generate realistic candidates.
Since their introduction, GANs have been used widely in several application domains e.g. video games, fashion modeling and astronomy. They are originally well-suited for unsupervised learning applications. However, they have been also successfully applied to (semi-)supervised learning and reinforcement learning. 

\begin{figure}[bht]
\centerline{\includegraphics[width=.5\textwidth]{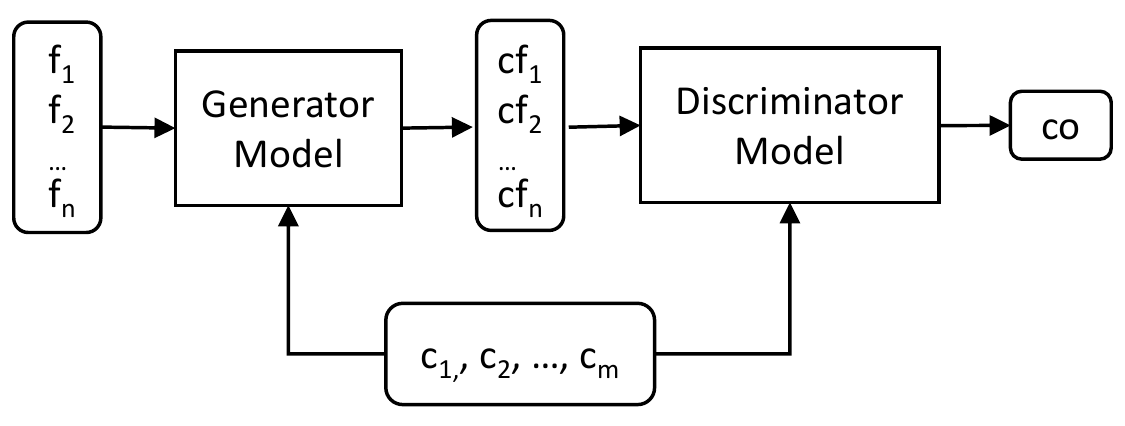}}

\caption{CGANs.}

\label{fig:cgan}
\end{figure}

Conditional GANs (CGANs) \cite{mirza2014conditional} are a type of GANs that can generate candidates conforming to some conditions. These conditions may be expressed as class labels, which make GANs applicable to supervised learning. Conditions are usually modeled by an additional linear layer in CGANs. The general architecture of a CGAN is presented in Figure \ref{fig:cgan}. In this figure, the generator model transforms each input vector $<f_1, f_2, ..., f_n>$ to a candidate vector $<cf_1, cf_2, ..., cf_n>$ considering the conditions specified in the condition vector $<c_1, c_2, ..., c_m>$. The discriminator model then generates a Boolean variable $co$ indicating whether the candidate vector conforms to the conditions in the condition vector.

Loss and accuracy are the metrics typically used to evaluate the performance of neural network models. However, these metrics are not applicable to GANs and their variants. In other words, we can not use these metrics to determine when to stop training a CGAN model.  \textit{Frechet joint distance} (FJD) \cite{devries2019evaluation} is a quantitative metric recently proposed to evaluate the performance of CGANs. This metric is an extension of the well-known \textit{Frechet inception distance} (FID) metric \cite{heusel2017gans} proposed for GANs. FJD and makes FID applicable to conditional generation settings.
\section{\method}\label{sec:method}
\begin{figure*}[htb]
\centerline{\includegraphics[width=\textwidth]{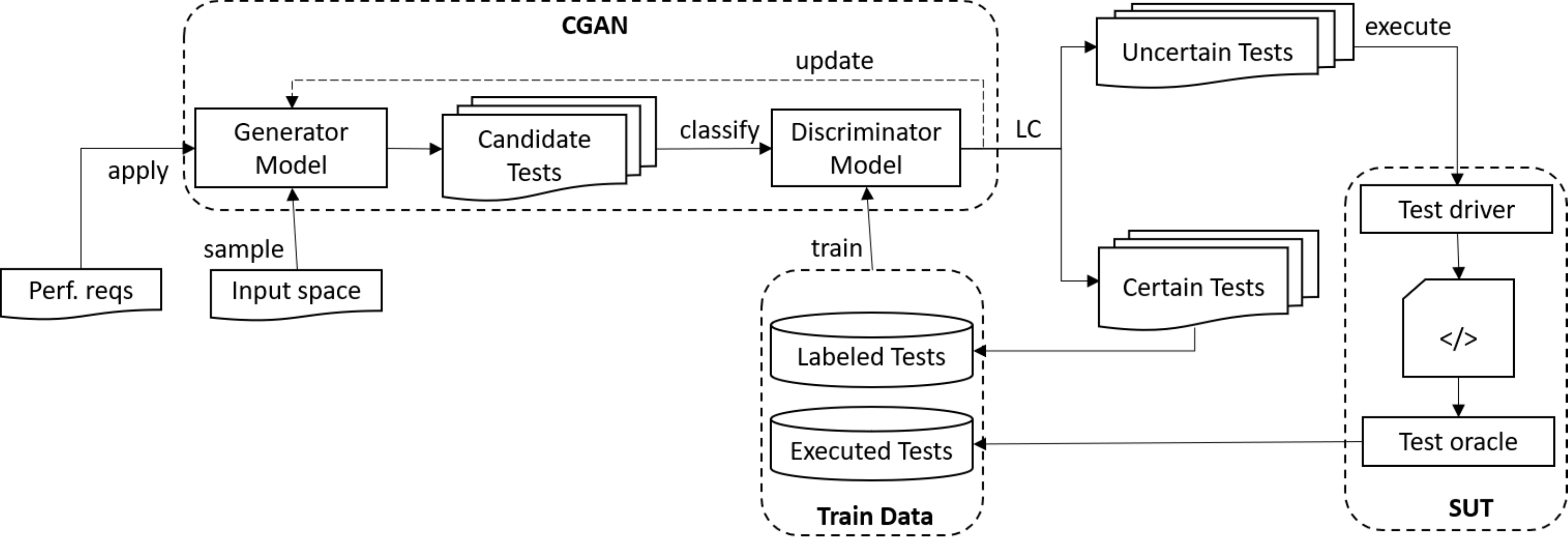}}

\caption{An overview of \method.}

\label{fig:method}
\end{figure*}

In this section, we elaborate the proposed ADL method for generating positive tests. An overview of \method is presented in Figure \ref{fig:method}. Accordingly, in each iteration new tests are generated by the generator model, which  submits them to the discriminator model for classification. The discriminator model is a binary classifier that tries to distinguish real tests from synthetic tests generated by the generator model. Each input of the discriminator model is labeled with a coding representing the performance requirement that the test targets. This label is embedded as a condition in the discriminator model (see Section \ref{sec:CGAN} for more details about conditions in CGANs). Indeed, a trained CGAN model can be easily used to generate tests that challenge a specific performance requirement.

In the next step, \method uses the LC measure (see Equation \ref{eq:lc}) to divide the classified tests into two groups based on the prediction confidence of the discriminator model. Certainly, the group with low confidence includes tests with high uncertainty about whether they are real and executed or synthetic. This group is then sent to the test driver for execution on the SUT. The group with high confidence together with the tests executed on the SUT are then exploited to fine-tune the discriminator model. By using both executed and labeled tests for training the discriminator model, not only its classification accuracy will be improved (through the knowledge learned by interacting with the SUT), but also it will learn more robust features through self-training \cite{tomanek2009semi}. 

\subsection{CGAN Architecture}\label{sec:CGAN_arch}
As mentioned above, \method relies on a CGAN model to learn about the SUT and generate appropriate performance tests to be executed on the SUT. Both the generator and discriminator models in this CGAN are fully-connected deep neural networks. In the following sub-sections we provide more details about the architecture of each of these networks and how they form a CGAN model.

\subsubsection{Generator Model}
The generator model is responsible for generating new tests by sampling from the input space i.e. a set of input variables including SUT's configuration parameters, method/service IDs and their input parameters. The input variables are represented as a tuple of integer or floating point numbers. The generator model takes as input random tests (formed by sampling from the input space), such that to each test a random label is assigned. Then it tries to transform them to tests that the discriminator model cannot distinguish from real tests. The architecture of the generator model consists of an embedding layer, two hidden layers and one output layer. The purpose of the embedding layer is to turn labels into dense vectors of fixed size 10. Each hidden layer is a fully-connected layer with 128 neurons and a \textit{ReLU} activation function. The output layer is also fully-connected and has a \textit{Tanh} activation function. 

\subsubsection{Discriminator Model}
The discriminator model is responsible for classifying the input tests to fake and real groups. The architecture of this model is similar to that of the generator model, except that it adds a flatten layer before the output layer and the output layer uses a \textit{Sigmoid} activation function. 

The discriminator model uses \textit{binary cross-entropy} as the loss function and \textit{Adam} optimizer with a learning rate of $10^{-4}$ for updating the network weights. Adam optimizer is an efficient extension of stochastic gradient decent specifically designed for deep neural networks.

\subsubsection{CGAN Formation}
The CGAN model can be defined as a combination of the generator model and the discriminator model such that the discriminator model is added on top of the generator model. The CGAN model is actually responsible for updating the weights of the generator model based on the classification errors made by the discriminator model.  So, the generator model does not update itself. The CGAN model also uses \textit{binary cross-entropy} as the loss function and \textit{Adam} optimizer with a learning rate of $10^{-4}$.

\subsection{Test Execution}
\method is fully automated and does not require any prior knowledge about the internals of SUT for test execution. Instead, it requires the performance requirements, the value domains of input test data and an API to call the test driver and get the test output from the test oracle. Only the tests that the discriminator model is uncertain about their fakeness, will be executed on the SUT during the CGAN training phase (see Section \ref{sec:train} for more details). This would reduce the testing effort in comparison to methods based on random testing and passive learning.

\subsection{CGAN Training}\label{sec:train}
Algorithm \ref{alg:ADLTraining} formalizes the CGAN training steps. In lines 1-5 the algorithm builds the CGAN model based on the architectures explained in Section \ref{sec:CGAN_arch}, and initializes the sets of labeled and executed tests to empty sets. The training loop of the algorithm is outlined in Lines 6-20. After each iteration, the FJD measure (see Section \ref{sec:CGAN}) is evaluated on the candidate tests synthesized by the generator, and the loop is repeated as long as the FJD value is higher than a given threshold. 

After training the discriminator, the generator model generates new tests by sampling from the input space and assigning a random label to each generated test. The new tests are then submitted to the discriminator model for classification. The classification errors made by the discriminator model are then used to update the weights of the generator model (line 11). Next, the classified tests are ranked based on the confidence in their most likely class (we have two classes: fake and real), and the ones with the least confidence are submitted to the test driver for execution on the SUT (lines 12-17). The maximum number of executed tests is determined by the test budget.

\begin{algorithm}
\SetAlgoLined
\KwIn{InputSpace, PerfReqs, TestBudget}
 Gen = new GeneratorModel(PerfReqs)\;
 Disc = new DiscriminatorModel(PerfReqs)\;
 CGAN = new CGAN(Gen, Disc)\;
 ExecutedTests = $\emptyset$\;
 LabeledTests = $\emptyset$\;
 \Do{$FJD(CandidateTests, ExecutedTests) > FJD\_Thr$}{
    UncertainTests = $\emptyset$\;
    Disc.train(ExecutedTests $\cup$ LabeledTests)\;
    CandidateTests = Gen.synthesize(InputSpace)\;
    ClassifiedTests = Disc.classify(CandidateTestS)\;
    CGAN.updateGen()\;
    \While{$|UncertainTests| < TestBudget$}{
        LowConfTest = findLC(ClassifiedTests)\;
        UncertainTests.add(LowConfTest)\;
    }
    TestDriver.execute(UncertainTests)\;
    ExecutedTests = ExecutedTests $\cup$ TestOracle.outcome()\;
    CertainTests = ClassifiedTests - UncertainTests\;
    LabeledTests = LabeledTests $\cup$ CertainTests\;
 }
 save(CGAN)\;
 \caption{AL algorithm for training the CGAN}
 \label{alg:ADLTraining}
\end{algorithm}

\subsection{Test Generation}\label{sec:tcgen}
After training the CGAN, the discriminator model can be discarded, and we can use the generator model to generate positive tests that violate a specific performance requirement. The test generation steps are elaborated in Algorithm \ref{alg:ADLGeneration}. This algorithm takes as input the target requirement and the desired size of the test suite. Then, it generates new tests iteratively by asking the generator model to make predictions with respect to the given requirement. It is possible that repetitive tests are generated since the generator model cannot guarantee the uniqueness of each generated test. Also, since the generator model is not 100\% accurate, there is no guarantee that all of the generated tests are positive. 

\begin{algorithm}
\SetAlgoLined
\KwIn{PerfReq, Size}
\KwOut{goodTCs}
 TestSuite = $\emptyset$\;
 CGAN = loadCGAN()\;
 Gen = CGAN.Gen\;
 \While{$|TestSuite| < Size$}{
    Test = Gen.predict(PerfReq)\;
    TestSuite.add(Test)\;
    
 }
 \caption{Algorithm for generating positive tests targeting a specific performance requirement}
 \label{alg:ADLGeneration}
\end{algorithm}
\section{Evaluation}\label{sec:exp}
To examine the practicality of \method and compare its performance and accuracy to other state-of-the-art methods, we implemented a prototype of \method and performed some experiments on a web application called RUBiS \cite{amza2002specification}. This application provides a core implementation for an auction website, and it has been widely used as a benchmark application for performance analysis in academia. In the following, we provide details about the implementation, the setup of the experiments and the outcomes. 

\subsection{Implementation}
\method is implemented using Python 3.6 and on top of the TensorFlow platform. TensorFlow is a popular ML platform supplying efficient implementations for various machine learning algorithms. To interact with the SUT, the Python program relies on an interface including a method for executing the tests generated by the CGAN model, a method for receiving the test results and recording them in the train dataset, and two methods for pre-processing and post-processing the test data. The pre-processing method transforms the SUT data to a format suitable for the CGAN model, and the post-processing method transforms the data generated by the CGAN model to the input format of the SUT. The implemented CGAN model expects floating point data in range $[-1, 1]$, which may differ from what expected by the SUT. Furthermore, the SUT may have categorical or integer inputs, which need to be converted to floating point values before being processed by the CGAN model. The categorical data (if any) need to be first converted to integers, and then to floating point.

\subsection{Experiment Setup}\label{sec:exp_setup}
To host the front-end of RUBiS, we used an Apache 2.4.46 web server with PHP 7.3.21, and MySQL 5.7.31 was also used as the back-end database. For performance analysis, we were interested to measure the elapsed execution time $t_exe$ on each executable test with a maximum threshold of 1 second. Since we have only one performance requirement, the tests processed by the CGAN model are labeled with either 1, meaning that their execution time is more than 1 second, or 0 otherwise. Apparently, we are interested to generate tests with label 1.
To make our results comparable with the recent work, we made the same assumptions as in \cite{ahmad2019exploratory}\cite{porres2020automatic}. We defined a test template as a sequence of the three HTTP requests listed in Listing \ref{lst:tc}. 

\begin{lstlisting}[frame=single, caption={TC template for  RUBiS},label={lst:tc},captionpos=b]  
/SearchItemsByRegion.php?category=CID &
categoryName=CN & region=RID
/ViewItem.php?itemId=IID
/ViewUserInfo.php?userId=UID

\end{lstlisting}

To become executable, this template requires values for four variables i.e. CID, RID, IID and UID. In \cite{ahmad2019exploratory}\cite{porres2020automatic}, the following  value ranges are considered for these variables: $[1, 20]$, $[1, 62]$, $[1, 50]$, and $[1, 50]$ respectively, which leads to 3,100,000 value combinations. Similar to \cite{ahmad2019exploratory}\cite{porres2020automatic}, we injected 20 uniform bottleneck clusters among these combinations, such that each bottleneck would increase the elapsed execution time of the corresponding test by five seconds. At the end, there were 292095 combinations with bottlenecks, which were labeled with 1. After preparing the dataset, we designed four experiments, one based on passive learning and three others based on AL. Table \ref{tab:experiments} presents the total number of tests, and the quantity of tests with bottlenecks (b-tests). In all of the experiments, the batch size for training the CGAN was 64, so the number of sampling steps per each epoch could be easily calculated for each experiment by dividing the dataset size by 64.

\begin{table}[htb]
    \centering
    \caption{Experiments}
    \begin{tabular}{| l | l | l | l |}
        \hline
         \textbf{Experiment} & \textbf{\# tests} & \textbf{\# b-tests} & \textbf{Epoch size}\\ \hline
         1 & 3,100,000 & 292095 & 48437\\ \hline
         2 & 1,000,000 & 73565 & 15372\\ \hline
         3 & 500,000 & 26133 & 7812\\ \hline
        4 & 100,000 & 7474 & 1562\\ \hline
    \end{tabular}
    
    \label{tab:experiments}
\end{table}

\subsection{Experiment 1: Passive Learning}\label{sec:passive}
\begin{figure}[bht]
\centerline{\includegraphics[width=.5\textwidth]{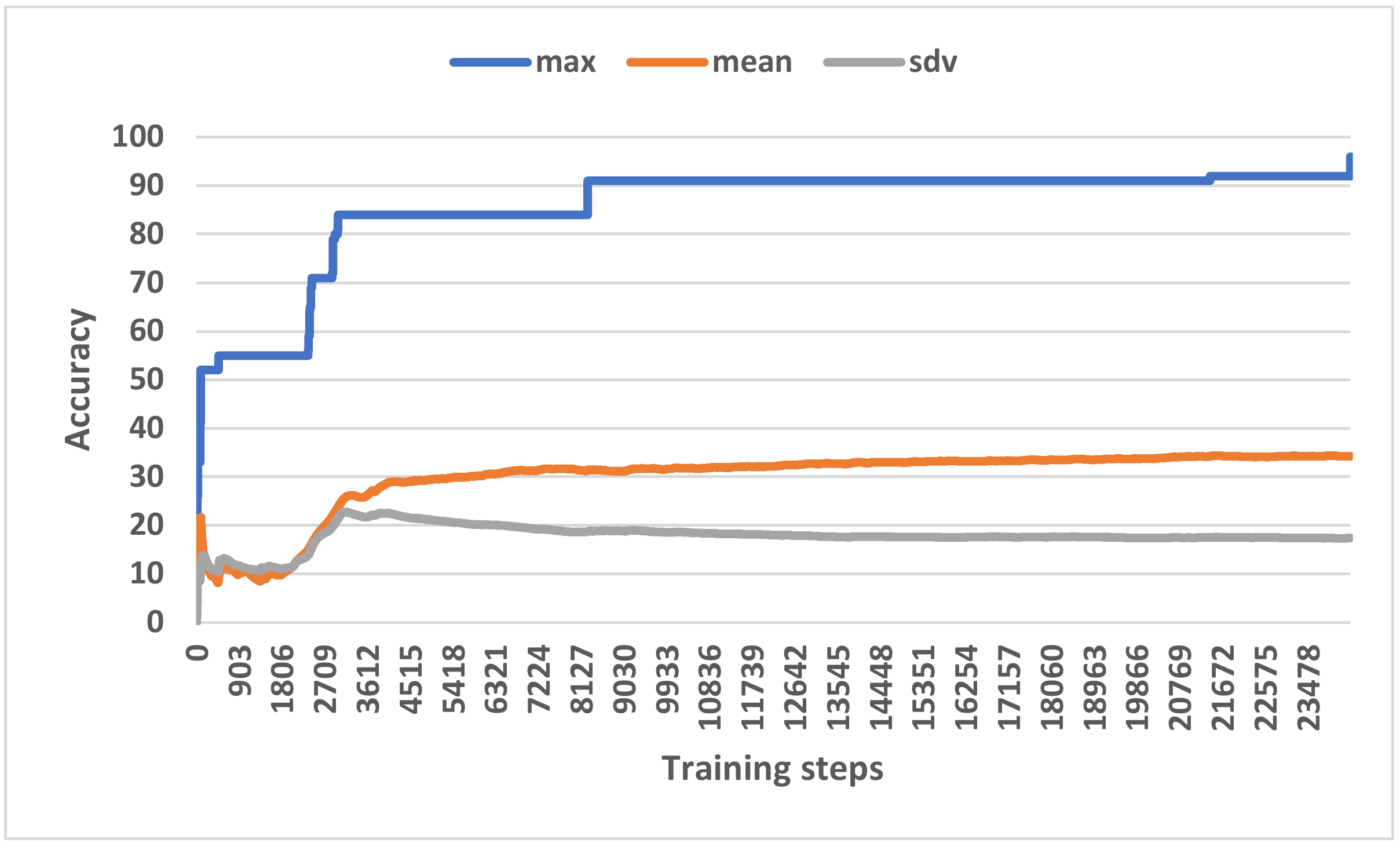}}

\caption{Accuracy of the CGAN in the first experiment.}

\label{fig:passive}
\end{figure}

In the first experiment we trained the CGAN model with the prepared dataset and analyzed its accuracy in generating tests that had bottlenecks (i.e. positive tests). 
In each training step, a batch of 64 tests was selected to train the discriminator model. This batch included 32 real and randomly selected tests from the input dataset and 32 fake tests generated by the generator model. For simplicity and intuitiveness, we did not use FJD to measure accuracy. Instead, we asked the generator model to synthesize 100 positive tests at the end of each training step, and evaluated the accuracy of the model by comparing the outputs with the b-tests stored in the dataset. The stopping criterion for the training process was reaching an accuracy of 96\%. Figure \ref{fig:passive} reports the number of generated positive tests with respect to the number of training steps. Accordingly, we could reach a model with 50\%, 80\% and 96\% accuracy in the first epoch and after 85, 2925 and 24377 training steps respectively. These results indicate that the CGAN model can learn the distribution of the injected bottleneck clusters fast. However, if we look at the values of \textit{mean} and \textit{standard deviation}, we notice significant changes in the accuracy of the CGAN model during the training process. But, as we proceed in the training process, the model becomes more stable and undergoes fewer updates.  

\subsection{Experiments 2-4: Active Learning}\label{sec:active}
In the AL experiments, only a portion of the dataset was available for the prototype implementation initially. For example, the first 1,000,000 records were available for the first experiment (see Table \ref{tab:experiments}). All of the AL experiments used the same implementation of Algorithm \ref{alg:ADLTraining}, and the batch size and accuracy measures were the same as experiment 1. However, in contrast to experiment 1, the tests generated by the generator model were not directly used to train the discriminator model. Instead, the discriminator model was first asked to predict the labels of those tests. These labels indicated whether the generated tests were real or synthetic. The labeled test were then sorted based on the certainty of the discriminator model about the assigned labels. Half of the tests with more uncertainty were then selected for active labeling based on the distribution of the bottleneck clusters. Active labeling was performed by a method which compared the generated input values with the value ranges of the bottleneck clusters in the whole dataset. The labeled tests were then used as real tests to update the discriminator model. 

\begin{figure*}[!t]
\centering
     \begin{subfigure}[b]{.49\textwidth}
         \centering
         \includegraphics[width=\textwidth]{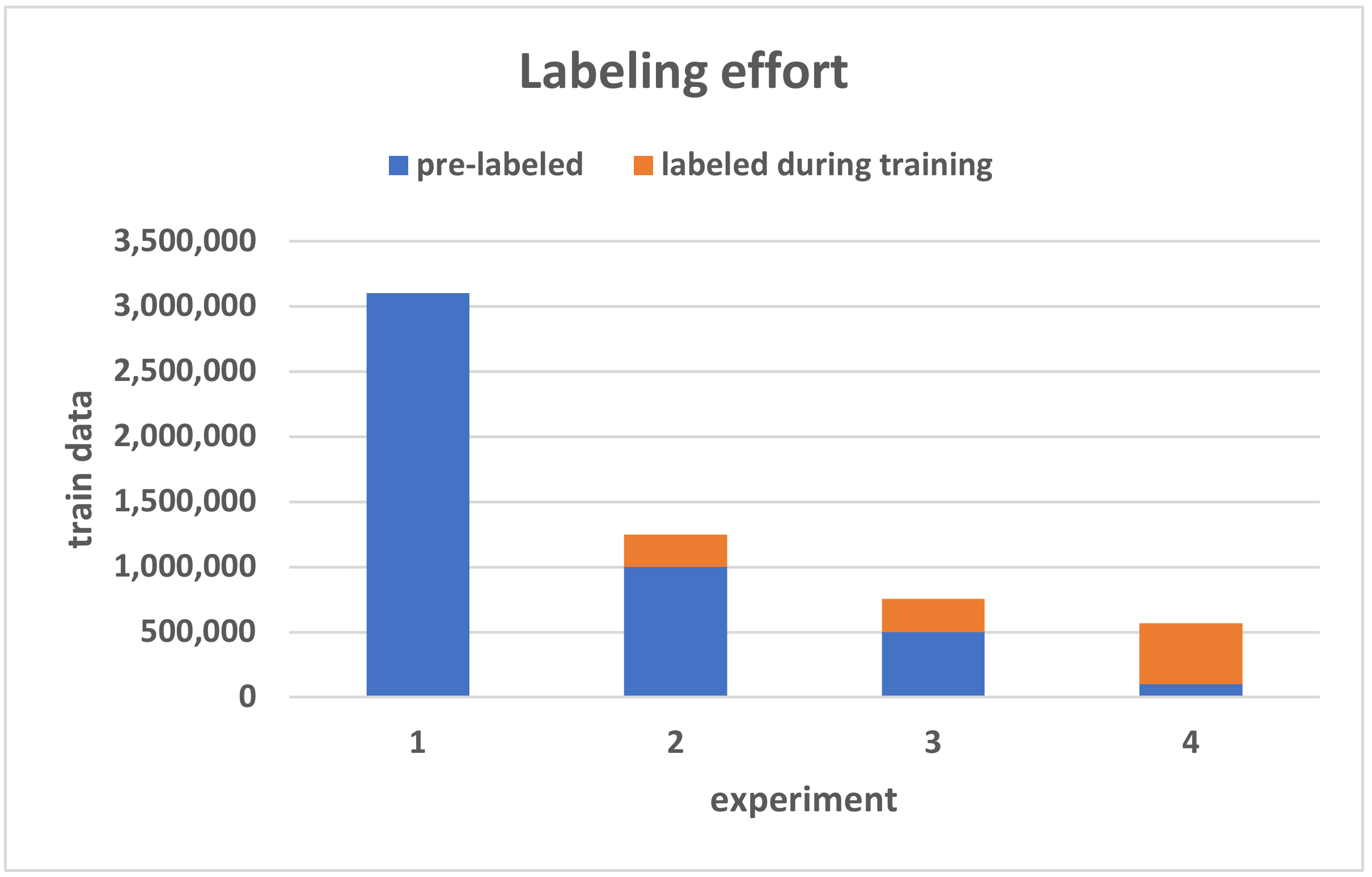}
         \caption{labeling effort}
         \label{fig:l_effort}
     \end{subfigure}
     \hfill
     \begin{subfigure}[b]{.49\textwidth}
         \centering
         \includegraphics[width=\textwidth]{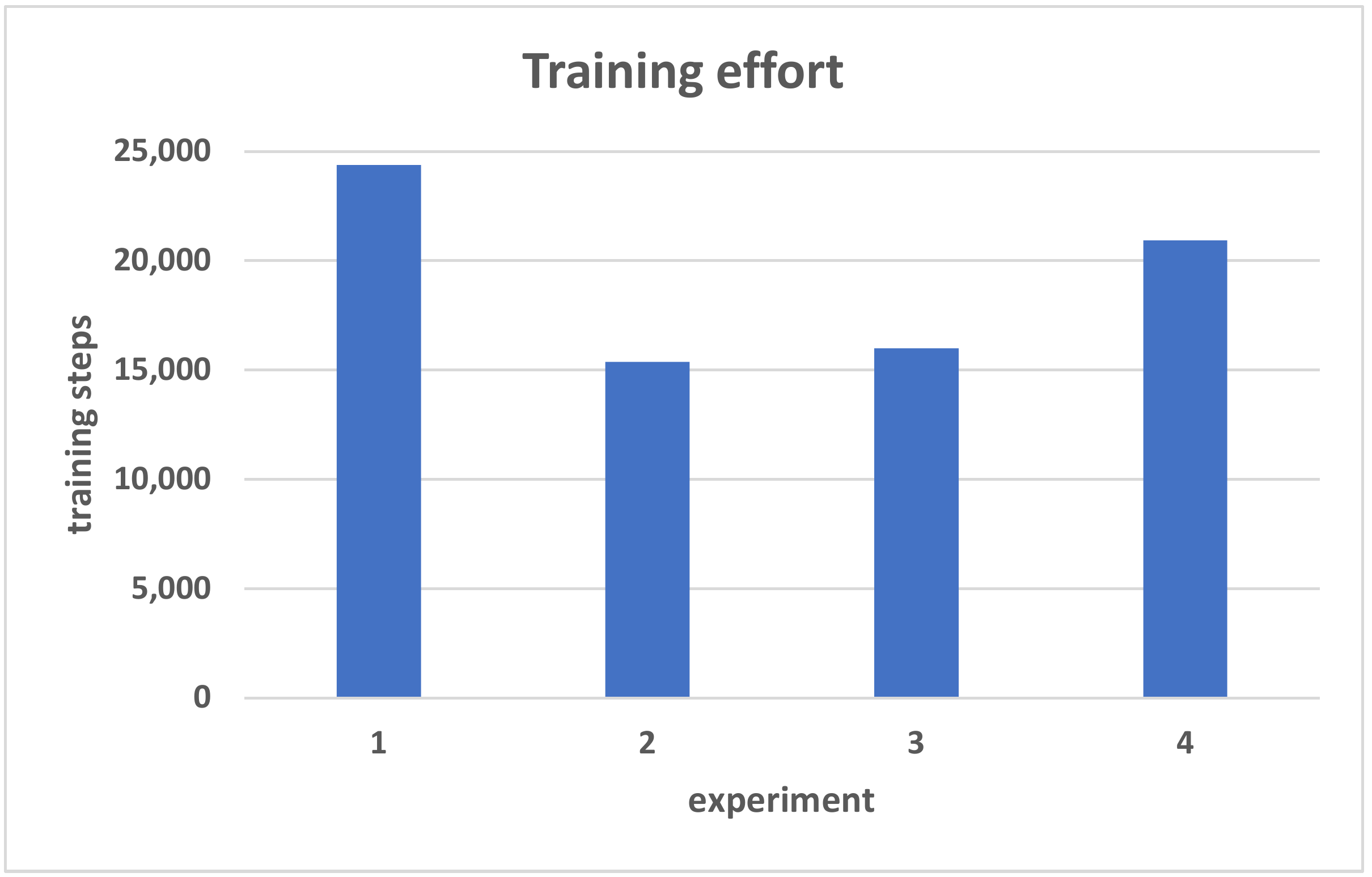}
         \caption{training effort}
         \label{fig:t_effort}
     \end{subfigure}
        
\caption{Effort comparison between the active and passive learning experiments.}

\label{fig:effort}
\end{figure*}

Figure \ref{fig:effort} provides a comparison between the labeling and training efforts of the four AL experiments discussed above. These results confirm that we can achieve a high accuracy with less labeling and even training effort in comparison to passive learning. However, since labeling is costly in several cases, it would be interesting to find the minimum size of the pre-labeled dataset suitable for an AL scenario to achieve a certain level of accuracy. As the results of experiment 4 indicate, if we have too few pre-labeled data, then we need to spend more labeling effort during the training process.

\subsection{Other Test Generation Methods}
In addition to the four experiments discussed above, we performed another experiment to compare the performance of \method with random testing and two recent ML methods i.e., PerfXRL \cite{ahmad2019exploratory} and DN \cite{porres2020automatic}. As detailed in Section \ref{sec:exp_setup}, we made the same assumptions as the ones made in \cite{ahmad2019exploratory}\cite{porres2020automatic} for doing experiments on the RUBiS application. In addition to these methods, we were interested to compare \method with random testing.
To perform random testing, we generated several combinations of input values by uniformly sampling from the domains of the input variables specified in Listing \ref{lst:tc}. Then, for each combination we verified whether it belonged to any of the bottleneck clusters or not. If yes, then the corresponding test would be a positive TC and negative otherwise. 

\begin{figure*}[!b]
\centerline{\includegraphics[width=.7\textwidth]{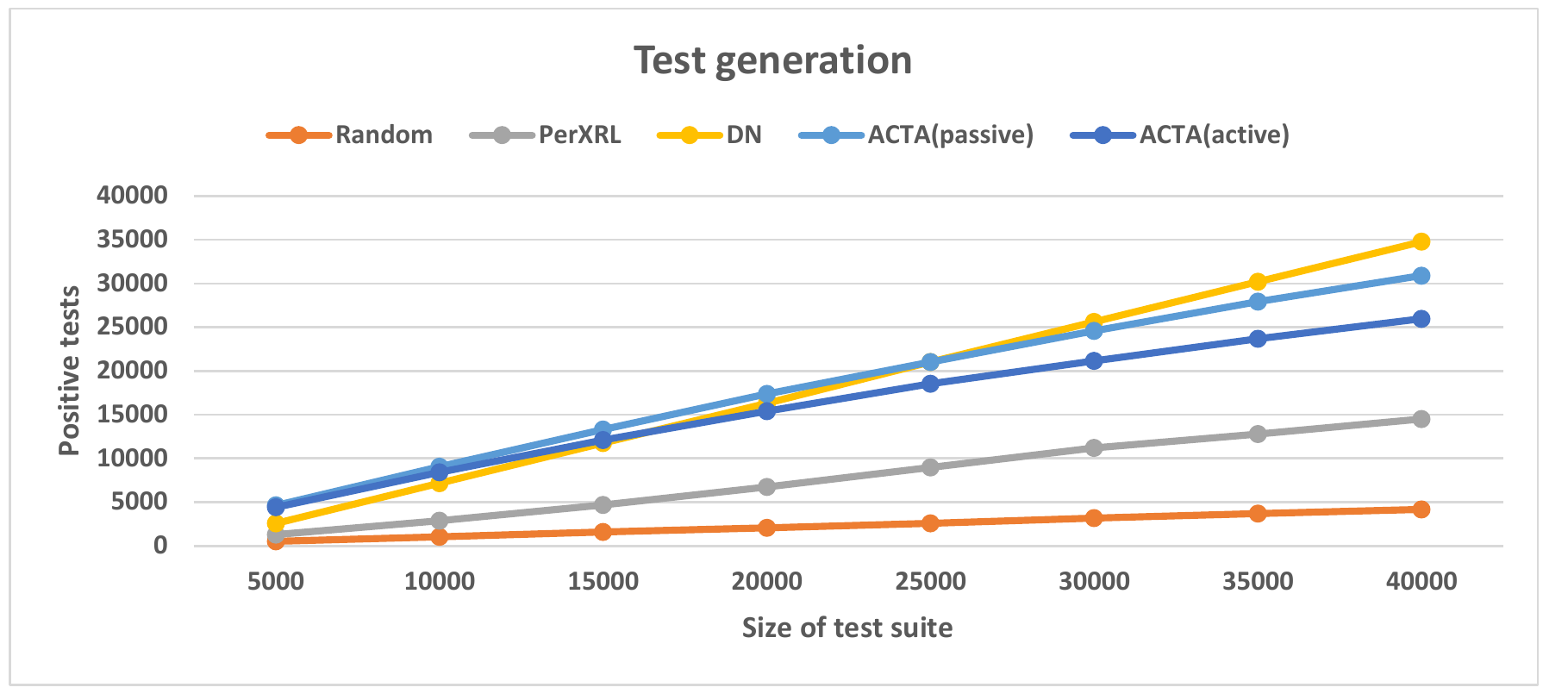}}

\caption{Comparison with other methods.}

\label{fig:rand}
\end{figure*}

Figure \ref{fig:rand} shows the relation between the number of positive tests and the total number of tests generated by \method, PerfXRL, DN and random testing. To compare \method with the other three methods, we used two saved CGAN models for test generation. The first CGAN (called pCGAN) was trained with passive learning in experiment 1, and the second CGAN (called aCGAN) was trained with active learning in experiment 2. As shown in Figure \ref{fig:rand}, the CGAN models performed better than random testing and PerfXRL in all cases. However, they performed worse than DN for large test suites. In particular, as we increased the size of the test suite, we noticed that the CGAN models tended to generate more duplicate tests. We also observed that aCGAN generated more duplicates than pCGAN which could be due to being trained with fewer real tests. As explained in Section \ref{sec:active}, the size of the pre-labeled dataset used in experimet 2 was 1,000,000 which is less than one third of the dataset used to train pCGAN. Since the bottleneck clusters were distributed uniformly across the initial dataset, the chance of finding positive tests in the training data would be also one third of pCGAN. However, the results indicate that the performance of aCGAN is close to pCGAN, which means that aCGAN can generate more diverse tests than pCGAN. These results confirm the assumption that letting the ML algorithm choose the data to train from may lead to a better performance.

\section{\method Applied to DevOps}\label{sec:discuss}
In a DevOps setting, the goal is to continuously deliver high quality software products. Here, small changes are merged rapidly with the baseline product and new versions need to be tested within a relatively short period of time. \method can be easily applied to a DevOps setting. In fact, the CGAN model trained by Algorithm \ref{alg:ADLTraining} can be updated to generate tests suitable for testing the performance of new versions of the SUT. The update procedure is described by Algorithm \ref{alg:ADLCI}. 

\begin{algorithm}
\SetAlgoLined
\KwIn{InputSpace, TestBudget}
 LabeledTests = $\emptyset$\;
 OldTCs = $\emptyset$\;
 \While{$|OldTCs| < TestBudget$}{
    OldTCs.add(ExecutedTCs.lastItem())\;
 }
 OldResults = OldTCs.results()\;
 TestDriver.execute(OldTCs)\;
 ExecutedTCs = TestOracle.outcome()\;
 \If{$OldResults \ne ExecutedTCs.resultes()$}{
    CGAN = loadCGAN()\;
    Gen = CGAN.Gen\;
    Disc = CGAN.Disc\;
    \Do{$FJD(ExecutedTCs, CandidateTCs) > FJD\_Thr$}{
        UncertainTCs = $\emptyset$\;
        Disc.train(ExecutedTCs $\cup$ LabeledTCs)\;
        CandidateTCs = Gen.synthesize(InputSpace)\;
        ClassifiedTCs = Disc.classify(CandidateTCS)\;
        CGAN.updateGen()\;
        \While{$|UncertainTCs| < TestBudget$}{
            LowConfTC = findLC(ClassifiedTCs)\;
            UncertainTCs.add(LowConfTC)\;
        }
        TestDriver.execute(UncertainTCs)\;
        ExecutedTCs = ExecutedTCs $\cup$ TestOracle.outcome()\;
        CertainTCs = ClassifiedTCs - UncertainTCs\;
        LabeledTCs = LabeledTCs $\cup$ CertainTCs\;
     }
     save(CGAN)\;
 }
 \caption{Algorithm for updating the CGAN in a DevOps setting}
 \label{alg:ADLCI}
\end{algorithm}

In this algorithm, first the set of labeled tests is invalidated (i.e. reinitialized to $\emptyset$) since the labels may not be valid for the new version. Then, the last \textit{TestBudget} tests executed on the previous version are reexecuted on the new version (lines 2-8). If the execution results were the same, we assume that the change impact on the performance of the SUT is low. In this case, Algorithm \ref{alg:ADLGeneration} can be used to generate positive tests for the new version using the old generator model. Otherwise, the discriminator model is trained with the newly executed tests and it will contest with the generator model until the FJD distance between the synthetic tests and the executed ones becomes low again (lines 13-28). The contest procedure is the same as that of Algorithm \ref{alg:ADLTraining}.
Certainly, if we have a low budget it is more probable that no change will be detected upon executing the old tests on the new version. In this case, we may continue to use the old generator model for  generating tests for the new version, which increases the probability of generating negative tests. Indeed, increase in the test budget would lead to a more updated generator model, hence more promising tests may be generated.

To see how this algorithm works in practice, we performed another experiment with the same setup as experiment 2 in Section \ref{sec:active}. In this experiment we removed 2, 3 and 4 bottleneck clusters in three steps, such that at the end 9 clusters were removed from the set of 20 clusters used in experiments 1-4. In each step we gradually updated a CGAN saved after experiment 2. We recorded the training effort spent in each step to reach the same accuracy as reported in Figure \ref{fig:rand} for experiment 2. If we had retrained the CGAN model, then we had to spend around 15,000 training steps. However, in this experiment the training effort to address the bottleneck updates were 1237, 1763 and 2429 training steps respectively. These results confirm that the CGAN model can relatively quickly adapt to minor changes in the test data.  

\section{Related Work}\label{sec:related}
The authors of \cite{chen2016generating} use symbolic execution to explore program paths and determine the probabilistic relation between the input data and the execution time of programs. The learned knowledge can then used to find input data causing the worst-case execution time. To overcome the low scalability of symbolic execution, the authors of \cite{koo2019pyse} propose to use reinforcement learning to learn policies from a small-scale test and apply them to a large-scale one. As another approach suitable for white-box testing, Lemieux et al. introduce PerfFuzz \cite{lemieux2018perffuzz} as a fuzzing tool for testing the performance of C programs. Starting from a set of random inputs, PerfFuzz iteratively generates inputs which exhibit worst-case algorithmic complexity in different components of the program. In each iteration, the inputs that execute a new program location (or increase code coverage) are saved to generate inputs in the next iteration. 

In \cite{moghadam2019machine,HelaliMoghadam6057}, a reinforcement learning framework is presented for testing the performance of software systems. In this framework, system state is defined in terms of the performance measures of interest, and the goal of the test agent is to learn a sequence of operations that lead to the highest degradation in those measures. As another work based on reinforcement learning, Ahmed et al. \cite{ahmad2019exploratory} introduce PerfXRL as an approach to find combinations of input data values that reveal performance defects in the SUT. In PerfXRL each combination of input values is represented as a state and any modification to that combination would change the system state. Here, the goal of the test agent is to find the set of states that trigger a performance bottleneck i.e. the execution time for the embedded values is more than a given threshold. This method is only applicable to discrete input spaces. However, in most cases we have continuous input variables to decide about. 

Shen et al. \cite{shen2015automating} combine the genetic algorithm with contrast data mining to find combinations of input values that make the SUT's execution time exceed a given threshold. In this approach, profilers are used to measure the execution time and access to source code is required to find methods that contribute to performance issues. Genetic algorithm is also used in \cite{ahmad2018identifying} for exploring the performance space of web applications and identifying user scenarios with the potential to create the highest level of utilization for a given system resource. In this work, Markov chains are used to model user scenarios, and the genetic algorithm finds the worst path in this model with respect to the utilization of the target resource. This path represents the worst user scenario. \method can also be used to identify high-utilization scenarios if we consider each decision variable in user scenarios as an input variable to explore for test generation.  

As a ML approach based on decision trees, in FOREPOST \cite{luo2017forepost, grechanik2012automatically} rules learned from execution traces are used to find input data that lead to intensive computations. The extracted rules describe the computational intensity of workloads in terms of the properties of the input data. In \cite{porres2020automatic} the authors propose to use DL for finding appropriate input combinations and generating good TCs automatically. The deep model is a discriminator model trained to be able to distinguish between positive and negative tests. Therefore, we still need to generate several random tests to find some positive ones. \method is also based on DL. However, \method is active and generative rather than being passive and discriminative.

\section{Concluding Remarks}\label{sec:conclusion}
Taking advantage of the recent advances in machine learning, we presented \method as a new method for automated performance test case generation. \method is suitable for black-box settings, where (1) we don not have access to source code and structural/behavioral models, (2) we deal with a potentially large input space consisting of variables of any categorical/numerical data type, (3) we have a limited test budget, and (4) the system is continuously changing.  
\method is implemented in Python and the experiments on a benchmark web application confirm its practicality and efficiency. Despite being originally proposed for performance testing, there is no limitation to use \method for testing other quality attributes e.g. dependability and security.   

In the current implementation, \method only supports basic active learning based on uncertain sampling. Providing support for other sampling strategies and more recent extensions of active learning (e.g. single-pass active learning \cite{lughofer2012single} and hybrid active learning \cite{lughofer2012hybrid}) are considered as future work. Furthermore, the architecture of the deep models in \method is simple. Investigating other types of layers (e.g. convolutional) and activation functions is also planned for future. Last but not the least, the input space of the benchmark application analyzed in this paper was rather small and comprised of quite a few variables. We still need to evaluate the scalability of \method to large input spaces with tens of variables.   

\section*{Acknowledgment}
This work was carried out during the tenure of an ERCIM ‘Alain Bensoussan’ Fellowship Program and has been partially supported and funded by the ITEA3 European \href{https://itea3.org/project/ivves.html}{IVVES} project. 

\bibliographystyle{plain}
\bibliography{refs}

\end{document}